# ANALYSIS OF THE EVOLUTION OF TERRESTRIAL RELIEF THROUGH INVERSE ELEVATION TRANSITION

ANÁLISIS DE LA EVOLUCIÓN DEL RELIEVE TERRESTRE SEGÚN LA TRANSICIÓN INVERSA DE COTAS

Yulisa Valverde-Romero[1], Milton Valverde-Romero[2], Jhair Airtor Tarazona-Valverde[3], Giorgio Aldair Tarazona-Valverde[4], Yehosúa Antony Tarazona-Valverde[5]

**RESUMEN**

Al analizar la Teoría de la Energía Sísmica y la Teoría de la Energía Volcánica entendemos que, como resultado de la sinergia de la energía interna con la energía externa en suelo propicio, se han ido produciendo cambios incalculables, evaluables en tempo geológico. Así, la presente investigación propone que la compleja evolución del relieve de la corteza terrestre es resultado de la transición inversa de cotas y la fluidez de las masas de agua de los océanos, como resultado, constantemente se van redefiniendo la ubicación, dimensión, forma y cantidad de los continentes, las islas y los océanos, comprendiendo que la geografía que observamos actualmente no es más que la foto del momento.

**Palabras clave**: relieve terrestre, continentes, océanos, islas, TES, TEV.

**ABSTRACT**

Through an examination of the Theory of Seismic Energy and the Theory of Volcanic Energy, we come to comprehend that the interplay of internal and external energies in conducive terrain has led to immeasurable transformations, assessable within geological time frames. Consequently, this study posits that the intricate evolution of Earth's crust relief stems from the inverse shift in elevations and the mobility of oceanic water masses. This, in turn, influences the location, size, shape, and quantity of continents, islands, and oceans. It is crucial to recognize that the geographical features we currently observe merely capture a momentary snapshot, due to the constant movement of the Earth's surface.

## Introduction

The Earth exhibits a distinctive topography. When observed with the naked eye, certain formations seem to arise from flat surfaces, while in other instances, they give the impression of emerging from other structures. The presentation of these formations often suggests a vertical force at play. Noteworthy contrasts further contribute to this uniqueness, such as volcanic strata in plains, the discovery of sediments and/or marine fossils on mountain summits, or remnants of civilizations at the ocean floor. Conversely, expansive aseismic zones convey an apparent tranquillity that contradicts the information gleaned from the analysis of their strata layers. These peculiarities hint at the existence of a terrestrial dynamic that, up to now, has proven challenging to interpret due to its seemingly ambiguous nature.

[1] Asociación CVR-TES Group. Huaraz, Perú. iD ORCID: 0000-0002-0428-2020. Yulisavalverde8@gmail.com
[2] Asociación CVR-TES Group. Huaraz, Perú. iD ORCID: 0000-0002-9860-2258. Miltonvalverde7@gmail.com
[3] Asociación CVR-TES Group. Huaraz, Perú. iD ORCID: 0000-0002-0090-0581. Jhairt79@gmail.com
[4] Asociación CVR-TES Group. Huaraz, Perú. iD ORCID: 0000-0003-0411-9714. Tarazonagiorgio12@gmail.com
[5] Asociación CVR-TES Group. Huaraz, Perú. iD ORCID: 0000-0003-3193-7181. yehosuatarazona51@gmail.cm

Indeed, through the application of analyses outlined by Claudio Valverde Ramírez in his Seismic Energy Theory –TES, by its acronym in Spanish– (Valverde & Valverde, 2020) and Volcanic Energy Theory[2] –TEV, by its acronym in Spanish– (Valverde & Valverde, 2021), foundational insights emerge for comprehending the Earth's relief evolution. Thus, emphasizing the initial framework articulated in TES and TEV theories, it is asserted that gaining a profound understanding of Earth's development necessitates a macroscopic perspective, perceiving it as a unified entity. This approach is crucial, because our dimension in relation to the Earth is minuscule, and if, despite understanding this difference, we insisted on proposing big studies mainly based on what we perceive, it would be equivalent to analysing the functioning of an organism solely on a macroscopic scale using information derived from a microscopic scale. Such an approach is expected to yield inaccurate results, as our existence and the planet's evolution unfold in distinct dimensions. Consequently, it is essential to seek equivalence in the research. Hence, for a clearer understanding of the aforementioned, we will commence with a proportionality analysis. In doing so, we will use the measurements of the most prominent land features, namely the highest mountain globally, Mount Everest (8,848 meters above sea level, equivalent to 8.8 km), and the deepest marine depression, the Mariana Trench (10,706 meters deep, equivalent to 10.7 km). These benchmarks, however, are deemed non-representative formations, as they do not even reach one-third of the Earth's crust average. Analysing it in relation to the Earth's radius, we realize that the dimensions of both the tallest mountain and the deepest trench are barely discernible, as the height of the highest mountain, for instance, represents only about 0.14% of the Earth's radius, and the depth of the deepest trench is approximately 0.17% of the Earth's radius.

Consequently, this preliminary analysis reveals that we inhabit an exceedingly thin and superficial layer of our planet. It's worth noting that despite this, our exploration and understanding of this thin membrane are limited. Moreover, when we consider our dimension in comparison to the geological formations mentioned earlier, it becomes evident that our existence is practically imperceptible within our environment. This initial analysis serves to sharpen our focus on the research, highlighting that not everything we perceive must be universally regarded as information, given that our existence unfolds in a dimension distinct from the environment in which we develop, and which is the object of study in the present investigation (planet Earth).

Subsequently, we will provide a brief overview of the referenced theories. TES (Valverde & Valverde, 2020) stands as the original and unpublished proposition elucidating the genesis and progression of seismic activity. It posits that this phenomenon arises from the intricate interplay between geomagnetism (internal energy) and solar photons (external energy) under specific favourable conditions, with climate and soil type playing pivotal roles (Valverde, et al., 2022; Valverde, et al., 2022b; Valverde, et al., 2023a). On the other hand, TEV (Valverde, et al., 2023b), building upon the foundation laid by TES, offers an insight into the formation and evolution of volcanoes. It approaches them in their actual dimensions on Earth, characterizing volcanoes as localized, superficial geo-combustions with a finite lifespan.

Therefore, upon examining the propositions presented in TES and TEV, we discern that both theories offer fundamental insights that serve as a foundation for comprehending various interconnected natural phenomena. These theories provide plausible and straightforward explanations for numerous geological mysteries that persist. In a scientific context, these answers significantly enhance our understanding of the intricate dynamics governing the development of our natural environment, specifically aiding in unravelling the complexities surrounding the evolution of terrestrial relief.

**Metodology**

The methodology utilized in this research is the analytical method, facilitating the detailed analysis of each element present on Earth's surface, as well as the identification of cause-effect relationships in geological times.

Beyond merely establishing formulas, it is essential to initiate the process with careful observation of the environment. It is imperative to perceive it in its true dimension and discern the relationships among various natural events and their coexistence.

**Formation of Mountains**

When addressing the alterations to the Earth's surface, it becomes crucial to initiate the examination with the formation of mountains, as per this study, they predominantly shape the diverse topography of the Earth.

Initially, according to the concepts proposed in TES, the convergence of external and internal energy triggers a sudden displacement of external energy due to their repulsive interaction. This displacement results in a vibrational motion characterized by a waning intensity. If this external energy is swiftly shifted across an area containing reactive materials, it also serves as activation energy, instigating geological combustion. Geological combustion is a chemical reaction that releases energy, leading to an elevation in temperature and the initiation of subsequent chemical reactions, including the release of gases, as explained by Valverde (Valverde et al., 2022a).

The emission of gases is, in part, a consequence of the combustion process, releasing carbon dioxide, along with the evaporation of liquid elements present in the region. Subsequently, with the concentration of gases and the expansion of participating elements in the combustion, there is an increase in pressure, resulting in a subsequent rise in temperature. This higher temperature, in turn, involves additional surrounding elements in the combustion process. As new components join the combustion, expansion continues, causing an escalation in gas production and pressure, leading to another temperature increase. This cycle, where the rising temperature involves previously resistant adjacent components, is termed geological chain combustion a process resembling an ascending spiral unfolding over geological time.

The geological chain reaction will undergo successive increases in pressure, externally evident through continuous expansion of the surface volume. These expansions can be observed as the initial phase in the creation of a mountain. As a consequence of these chemical reactions, the active materials accumulate, forming a cavity that gradually develops into the volcanic chamber. This progression is externally visible as the evolution of a mountain. When the cavity reaches its maximum capacity, a conduit is formed, referred to as a volcanic fistula (commonly known as a chimney). Through this conduit, the pressure from the volcanic chamber is released, ultimately finding its way through the thinnest or most fragile surface, typically the upper part. This process leads to a volcanic eruption, during which various products of geological combustion, such as volcanic lava, gases, ashes, etc., are released.

Examining this preliminary context, we comprehend that not every volcanic chamber expands until a portion of the active content is expelled, meaning not every mountain evolves to the point of external manifestation as a volcano. The reason for this lies in the fact that as long as fresh reactive components continue to be introduced into geological combustion, the volcanic chamber's volume steadily grows, leading to the construction of mountains. However, if the reactive materials are depleted before the cavity surpasses its containment capacity, the combustion gradually decelerates until extinguished, with no release of the volcanic chamber's contents.

Neither is it accurate to assume that only the tallest mountains experience volcanic activity. Various conditions can lead to mountain eruptions, and some analysed assumptions include:

a) When there's a substantial amount of reactive material in the area, geological combustion persists even after exceeding the volcanic chamber's containment capacity. This typically results in high-altitude volcanoes with prolonged emissions.

b) When the materials involved in geological combustion are highly reactive, the process becomes more active, preventing the volcanic chamber's cover from adequately expanding. This accelerates the formation of a conduit, releasing accumulated pressure. In such cases, the mountain may not attain significant height, the mountain structure might be less complex, and premature extinction of volcanic combustion is possible if not sufficiently consolidated.

c) An ample supply of liquid elements in geological combustion leads to increased water vapor, raising pressure and temperature and accelerating the chain reaction, resulting in precipitated volcanic activity. Similarly, these volcanoes are expected to be of lower altitude, with simpler mountain structures, possibly not well-consolidated geological combustion (leading to premature extinction), and predominantly short volcanic activity.

Since these reactions unfold over geological time, they are perceived as exceedingly slow processes beyond our typical life dimensions.

**Overview of the Phases in Mountain Development.**

Based on the above analysis, it can be stated that the evolution of mountains undergoes five primary stages, each with its designated name and description (Figure 1):

**1. Germinal:**

This marks the initial phase of geological combustion. At this stage, due to chain oxidation, the combustion product begins to collect, forming an incipient volcanic chamber, with a slight increase in surface volume.

**2. Evolution:**

Representing the most active period in geological time, this stage witnesses the consolidation and external manifestation of geological chain combustion. The surface volume progressively increases, determining the dimensions of the mountains based on soil type and reactive components present in the area.

**3. Summit:**

The maximum height achieved by a developing mountain characterizes this stage. It denotes the pinnacle of mountain development, from which they begin to regress, either because the components that act as fuel are exhausted and the cessation of geological combustion begins or because volcanic activity began where pressure is released. This is the reason why volcanic eruptions are heralded by a rise in pressure, leading to the expansion of the mountain structure (Patanè, et al., 2003). Consequently, after the eruption and release of pressure, it is expected that there will be a decrease in the height of the now volcanic mountain. However, because some volcanic activities occur at intervals of hundreds to thousands of years, it is possible that a coating forms in the crater, resistant enough to once again contain the pressure generated in the combustion, thereby allowing a new evolution of the mountain, reaching a second development and a new summit.

**4. Involution:**

This phase signals the decline of the mountain.

As the accumulation of reactive material is depleted, the extinction of volcanic combustion begins, causing a gradual decrease in the mountain's size.

Slow cooling and the absence of pressure contribute to the contraction of volcanic chamber contents, leading to a progressive reduction in mountainous relief.

**5. Extinction:**

The opposite of the germinal stage, extinction occurs when geological combustion is exhausted. It's crucial to bear in mind that the formation of mountains occurs across thousands of years, meaning that a seemingly brief period of inactivity does not necessarily imply the cessation of geological combustion. To illustrate this, we can examine the case of the Popocatepetl volcano in Mexico, with an estimated age of around 730,000 years (Macías, 2005). Over the last 20,000 years, it has undergone four significant explosive events at intervals of 14,000, 5,000, 2,150, and 1,100 years ago (Siebe, et al., 2006).

The mountain descends until crowding ceases, preserving an inconspicuous volume.

In the case of a volcano with extensive volcanic emissions, the level at extinction may be at or even below the plain's level.

Over time, erosion and other environmental factors play a definitive role, rendering the mountain barely recognizable and leaving behind its once imposing size. If a new mountain emerges nearby, its reduced volume may eventually vanish in the topography, covered by lava and other volcanic emissions.

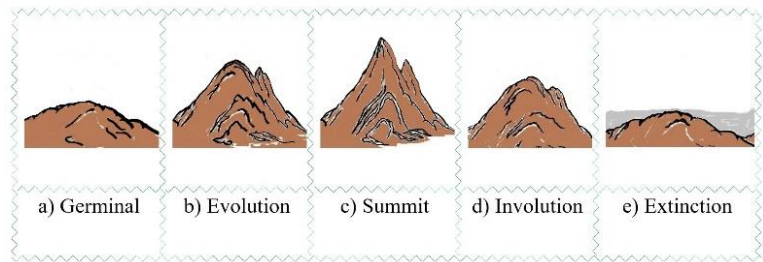

Figure 1: Representation of the development of the mountains in their respective stages.

During the germinal, evolution, and summit phases, heightened seismic activity can be anticipated in the region due to the conducive nature of the soil, acting as a good energy conductor. Additionally, the release of energy in combustion contributes to seismic waves, intensifying seismic vibrations. In contrast, during the stages of involution, particularly in advanced phases, and the extinction of the mountain, seismic activities are typically absent. This absence results from the cessation of chemical reactions, rendering the combustion residues poor conductors of energy. Consequently, seismic waves from nearby epicentres rapidly diminish, creating areas with minimal or no seismic activity.

Considering factors such as dimension, age, the period of involution, and aseismicity, the Ural Mountains serve as a representative case in the late involution stage. Despite being considered one of the oldest mountain ranges on Earth (250 to 300 million years old), these chains are still in the involution stage, and the height they could have reached during their summit stage is unimaginable.

In a nutshell, in a hypothetical examination, one can draw parallels between the evolution of the Earth's surface and the baking of a cake at various stages. Unlike the cake batter, the Earth's surface mass lacks a homogeneously distributed composition, and the internal energy does not uniformly cover the entire Earth's surface evenly. Contrasting characteristics to the even distribution and heating process of a cake. Although cake batter is homogenized before exposure to uniform heat during baking, the resulting cake surface is irregular. The rise in temperature triggers chemical reactions, releasing carbon dioxide ($CO_2$) and imparting volume to the product. Consequently, as the baking advances, surface elevations, eruptions, cracks, and even flat areas (depending on dough components, humidity, and heat exposure) emerge, similar to specific features on the Earth's surface.

In the post-cooking stage, as cooling proceeds, a loss of volume occurs, influenced by dough ingredients, cooking temperature, dough density, and the speed of temperature change. This resembles the process observed in the evolution-extinction phases of mountain development. Furthermore, the planet's evolution, driven by the interplay of internal and external energy, has occurred in periods and on a surface marked by clustered reactions. While acknowledging the complexity of both processes, this comparative analysis aids our understanding through analogy. Lastly, it is important to note that the initial presentation of this analogy as a hypothetical analysis, stems from the inherent complexity of recreating a cake baked at different times with non-homogeneous dough—a complexity comparable to that of the Earth's crust.

**Essential Involvement of Internal Energy in the Creation of Mountains and Other Geomorphologies.**

We will commence with the reference analysis conducted by Valverde (Valverde & Valverde, 2020), utilizing Hermetic philosophy as a framework. Principle of correspondence by Hermetic philosophy "As above, so below; as below, so above", encapsulates the idea that there are realms beyond our comprehension. By employing the principle of correspondence, we gain insights into aspects that would remain elusive otherwise. This universal principle finds application and manifestation across various domains (Los Tres Iniciados, 1990).

Drawing on the concept of unity, he established a correspondence between the Earth and an atom, noting that electrons orbit the atom in a manner similar to how internal energy circulates around the Earth. Recent studies indicate that electrons in atoms do not adhere to predictable paths; instead, their behaviour is characterized by probability distributions termed orbitals. Similarly, internal energy exhibits continuous movement without a fixed trajectory, with changes measured over hundreds to thousands of years (geological time). From our perspective (observational scale of months to years), we perceive a movement in circuits with a certain degree of regularity, sufficient for activating the reactive components in a given area.

Hence, we comprehend that over billions of years, internal energy has traversed the entire Earth's surface multiple times, contributing to the current topography. Consequently, evidence of this dynamism can be found across the Earth's surface at various points in time, supporting (Judson et al., 1981) assertion that even the flattest terrains were once part of ancient mountain.

Moreover, as internal energy activates reactive components along its path, favourable terrain results in the formation of reliefs whose dimension and dynamism depend on the quantity and reaction capacity of geological components in the area. Whether reactive material exists in a concentrated cluster giving rise to a solitary mountain, or in a greater longitudinal or dispersed concentration leading to the formation of a mountain range, or even if the material is agglomerated, activating components and creating irregular elevations as seen in massifs—each scenario is a consequence of the internal energy's interaction with geological components. It's worth noting that it has been observed that the influence of solstices and equinoxes leads to minor variations in the trajectory of internal energy (as will be explained in an upcoming publication). This results in subtle, temporary deviations in the path of internal energy, creating the possibility for mountain formation over a broader expanse. Alternatively, it is also conceivable that parallel lines may emerge, as seen in the Andes mountain range, where, in specific sections, mountain chains have developed in parallel.

Also, if there is an area with extinct mountains, and circumstances lead to the preservation of a cluster of reactive material, a new geological combustion may initiate when internal energy displaces through the region again. Typically, this new combustion is of smaller magnitude compared to the preceding one, but it does not preclude the development of volcanic activity. This scenario is observed in neo-residual volcanoes (satellite volcanoes) where a new combustion occurs amidst the remnants of a prior volcanic activity.

As a consequence of these reactions occurring at different geological times, mountains with heterogeneous characteristics can be identified. For instance, Mount Vesuvius in Italy is a neo-residual volcano of Mount Somma, where the body of Mount Vesuvius has developed within a portion of the ancient Mount Somma's crater. Another example is the Archède Mountains in the south of France, exhibiting significant volcanic formations dating from the Miocene to the Upper Pleistocene (Raynal, et al., 2023).

In short, from what is proposed in TES and TEV we understand that due to the displacements of the internal energy circuits, the reactive components of the Earth's surface have been activated, consequently, they have been consumed (losing their reaction capacity), therefore, seismic-volcanic activities in the past have been of greater intensity (information found through the study of various strata), so, currently, in the path of internal energy we observe limited volcanic activity in addition to the existence of aseismic areas, analysing that, in a still distant future, these seismic-volcanic events will have decreased even more in frequency and intensity.

**Inverse elevation transition.**

Inverse elevation transition or TIC (by its acronym in Spanish: *Transición Inversa de Cotas*), is the variation of the altitude of the reliefs. Due to the interplay of internal energy with external energy in favorable terrain, mountains emerge in specific areas of the planet. Conversely, when geological combustion concludes, as discussed earlier, the mountain undergoes a gradual reduction in volume (as detailed in the aforementioned stages). We refer to this asynchronous height variation resulting from the evolution and involution of mountains as the inverse elevation transition, recognizing that these geological events manifest irregularly and over geological time.

Considering the dynamic nature of the inverse elevation transition, one can envision that over millions of years, certain valleys presently known to us might transform into high-altitude mountains, possibly even exhibiting volcanic activity. Conversely, numerous grand mountains currently shaping our landscape may abate into plains or become part of marine relief. Alternatively, it's conceivable that some areas on the seabed could rise dramatically, comparable in height to Everest, as illustrated in Figure 2.

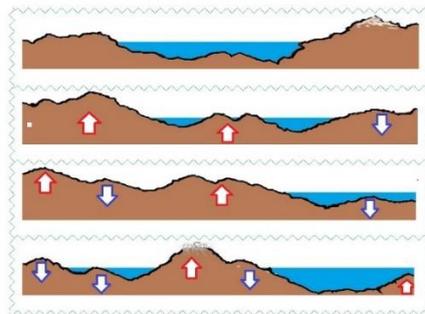

Figure 2: Illustration of the inverse elevation transition over geological time. A secondary effect observed is the continuous displacement of water masses influenced by gravity.

In a hypothetical scenario, we aim to visualize the Earth's surface without the presence of liquid elements, devoid of its covering biota, and stripped of recent sedimentary layers. Applying this filter may reveal a diverse yet relatively uniform structure, where specific distinctions between marine and terrestrial reliefs may not be readily apparent. It is noteworthy that earthquake and volcanic dynamics exhibit similar manifestations in both environments. Thus, our recognition of a terrestrial area having been part of the seabed in the distant past relies on evidence such as marine fossils or sedimentary layers in its strata, rather than discernible differences in geological structure.

As a result of this research proposal, we contend that the movements and displacements described in various geological studies are consequences of the inverse elevation transition. Consequently, the inverse transition of elevations not only provides answers to numerous questions but also naturally allows us to define the evolution of continents, islands, and oceans, as detailed below.

**Evolution of the Earth's surface and the shaping of continents, islands, and oceans.**

Due to the inverse elevation transition, while on the one hand there are mountains that evolve, others undergo involution, causing the water of the oceans, because of its fluidity, to constantly seek to level itself based on gravity. Thus, as a consequence of the fact that the inverse elevation transition is tardy but constant, as a reflex action, the water masses of the oceans continually move and thereby lead to the geological reestructuration, altering the locations, shapes, extensions, and numbers of oceans, islands, and continents, as depicted in Figures 2, 3, and 4.

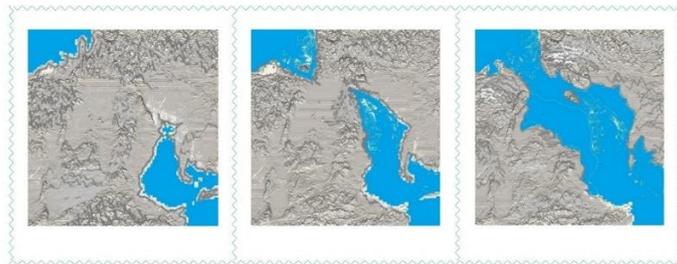

Figure 3: Displacement of water masses as a secondary effect of the inverse elevation transition. In this scenario, water slowly enters a land block, externally dividing geographic areas.

To delve into specifics, the inverse elevation transition and the fluidity of water masses play pivotal roles in shaping terrestrial relief. For instance, during the involution-extinction stage of a mountain within the interior of a continent or island, a depression forms, potentially reaching heights below sea level. It is anticipated that ocean water eventually move inland, and if it reaches the opposite coastline, it appear as a division of the land block, numerically increasing the areas not covered by water, as depicted in Figure 3. In an inverse scenario, with the inverse elevation transition, activating a large-scale geological combustion on the sea floor might lead to the emergence of a sea mountain sufficient to displace sea water between the emerging mountain and adjacent coastal mountains. This visualizes the emerging mountain as a new structure connecting land blocks previously separated by water masses, thereby reducing the number of islands or continents, as seen in Figure 4. Conversely, in the face of the inverse elevation transition, mountains may regress or evolve while maintaining their location, whether terrestrial or marine. This does not impact the quantity or shape of continents and islands but rather influences the relief.

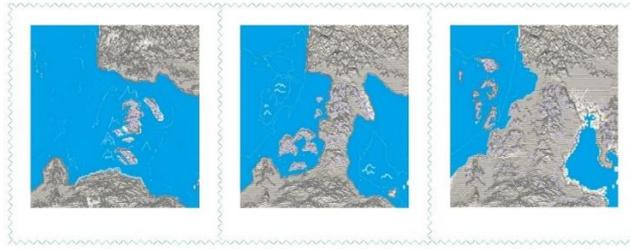

Figure 4: Depiction of mountain evolution, wherein such mountains may develop to the extent that the ocean water between the emerging mountain and the adjacent one is displaced. This results in visualizing it as a new structure connecting two areas that were superficially separated by bodies of water.

It's essential to emphasize that the displacement of water masses as a secondary effect of the inverse elevation transition occurs over thousands of years —an ample duration for the development of ecosystems to adapt to new environmental conditions. The development of this entire ecosystem, combined with recent sediments, whether marine or terrestrial, forms a thin film covering the Earth's surface. Therefore, based on this crustal covering, we may visually perceive islands and continents as structures distinct from the ocean floor.

Hence, the transformation of terrestrial relief due to the inverse elevation transition, accompanied by the displacement of water masses over time, has left numerous pieces of evidence. Notable examples include the case of Doggerland, where water entered at various periods—abruptly, slowly, and at times almost imperceptibly—eventually separating the United Kingdom from Europe (Walker et al., 2022). Another distinctive case of diverse evolution — otherwise almost impossible to interpret due to the variety of findings— is observed in the discoveries on the Tibetan plateau, characterized by a complex geological history with marine and continental remains, as well as evidence of volcanic activity. The plateau exhibits periods of evolution spanning millions of years, along with clearly differentiated biotic remains in different regions, leading to proposals regarding the ancient existence of microcontinents or islands (Xiumian, et al., 2022; Jian-Jun, et al., 2015). Additionally, findings such as archaic and juvenile cortical sources in Sweden (Andersson, et al., 2011), continental rocks in the ocean suggesting the past existence of submerged microcontinents in the Atlantic Ocean (Bortnikov, et al., 2022), and investigations into the presence of young and old rocks in various areas contribute to our understanding (Skolotnev, et al., 2010). Submerged ruins in different locations and depths of the ocean, such as Canopus and Herakleion in Egypt (Stanley, et al., 2001; Stanlet, et al., 2004) and Pavlopetri on the coasts of Greece (Henderson, et al., 2011). Also the identification of Zealandia as a geological continent (Mortiner, et al., 2017), among countless other examples.

**Conclusions:**

At times, we tend to perceive certain aspects as universal facts without fully considering the minuscule scale of our dimension compared to the vast body we seek to understand. It's crucial to acknowledge that our species' experiences are minimal, and our evolution can be seen as finite in the context of infinite development. Consequently, much of what appears regular, stable, or almost perpetual, when analysed in reference to the dimension of the body under study (in this case, planet Earth), may prove to be dynamic. What we currently know is likely just a snapshot of a particular stage or cycle. Taking a objective perspective, it becomes plausible to think that our understanding of Earth and the universe is not as constant as it may seem.

In this framework, an analysis of planet Earth aligns with Claudio Valverde Ramírez's theories, TES and TEV. When internal energy converges with external energy in favourable terrain, it initiates immeasurable changes over geological time. As long as clusters of reactive material persist, dynamism continues at the Earth's surface, leading to variations in relief and the shifting of water masses. This understanding suggests that continents and islands, will persistently undergo changes in location, shape, extension, and even number. These are not fixed structures, but integral components of a dynamic relief currently not submerged, rendering the most up-to-date cartography merely a snapshot in time.

**Acknowledgment:**

In memory of Claudio Valverde Ramírez; father, grandfather, and mentor; who became our guiding light illuminating our journey in science. Hugs to eternity.

**Funding:**

The work was conducted without funding.

**Competing interests:**

The authors declare no conflicts of interest.

**Author contributions:**

Conceptualization: YVR, MVR

Investigation: YVR, MVR, JATV, GATV, YATV

Writing – review & editing: YVR, MVR, JATV, GATV, YATV